\pdfoutput=1
\documentclass[11pt]{article}
\usepackage[utf8]{inputenc}
\usepackage[a4paper,margin=2.25cm]{geometry}
\usepackage[expansion=false]{microtype}
\usepackage{setspace}
\usepackage{amsmath,amssymb,bm}
\usepackage{graphicx}
\usepackage{booktabs,tabularx,longtable,array}
\usepackage{pdflscape}
\usepackage{caption}
\usepackage{subcaption}
\usepackage{enumitem}
\usepackage[round,sort&compress]{natbib}
\usepackage{xcolor}
\usepackage[hidelinks]{hyperref}

\graphicspath{{figures/}}
\hypersetup{
  pdftitle={Beyond the Four-Decade DIIS Default: Auxiliary-Curvature Acceleration of Self-Consistent-Field Calculations},
  pdfauthor={Peng Bao}
}

\newcommand{\aurora}{AURORA--SCF}
\newcommand{\Eh}{E_{\mathrm h}}
\newcommand{\A}{\mathrm A}
\newcommand{\D}{\mathrm D}

\title{\textbf{Beyond the Four-Decade DIIS Default:}\\
Auxiliary-Curvature Acceleration of Self-Consistent-Field Calculations}

\author{
	Peng Bao$^{1,*}$\thanks{Corresponding author: \texttt{baopeng@iccas.ac.cn}}
	\and
	Qiang Shi$^{1,2}$\
}

\begin{document}
\maketitle

\footnotetext[1]{Beijing National Laboratory for Molecular Sciences,
	State Key Laboratory for Structural Chemistry of Unstable and
	Stable Species, CAS Research/Education Center for Excellence
	in Molecular Sciences, Institute of Chemistry, Chinese Academy
	of Sciences, Zhongguancun, Beijing 100190, China}
\footnotetext[2]{University of Chinese Academy of Sciences, Beijing 100049, China}

\begin{abstract}
Direct inversion in the iterative subspace (DIIS), introduced in 1980, remains the practical default for accelerating Hartree--Fock and Kohn--Sham self-consistent-field (SCF) calculations. Many later methods improve robustness or iteration count, but the near-zero overhead of DIIS makes a broad reduction in total wall time difficult. We introduce \aurora{} (auxiliary-curvature unified Riemannian orbital-response acceleration), which evaluates energy and gradient only with the requested target Hamiltonian while obtaining most orbital curvature from an independent valence-STO-3G model. A target-level, transported L-BFGS history corrects the model mismatch; a shifted matrix-free solve, trust region, and geodesic orbital update control the step. In 16 direct CPU RHF pairs spanning 137--1484 atomic orbitals, \aurora{} was faster in every case, reducing mean wall time by 26.2\% and target J/K builds by 33.3\%. In a separate set of 63 converged, energy-matched density-fitted GPU pairs spanning seven molecules, closed- and open-shell formalisms, and HF, PBE, B3LYP, and M06-2X, it was again faster in every included pair, with mean reductions of 26.5\% in wall time and 31.9\% in target J/K builds. Focused direct CPU and GPU sweeps show mean wall-time reductions of 30--34\%. These results establish a specific advance beyond the four-decade DIIS default: transferred, secant-corrected curvature can reduce total SCF wall time, not merely iteration count, without changing the target stationary equations. The same optimization pattern also suggests a route to faster orbital optimization elsewhere in quantum chemistry and to multifidelity optimization across scientific computing.
\end{abstract}

\section{Introduction}

The SCF bottleneck is unusually resistant to optimization. Each useful outer iteration requires a target-basis Fock build--Coulomb and exchange for HF, plus exchange--correlation evaluation for Kohn--Sham DFT--whereas a DIIS extrapolation costs almost nothing. Pulay's 1980 and 1982 algorithms therefore established a formidable practical baseline: a method with imperfect global behavior but excellent local efficiency and negligible overhead \citep{pulay1980,pulay1982}. More than four decades later, DIIS or a DIIS-centered hybrid remains the default in many production workflows.

SCF optimization has not stood still. Quadratically convergent SCF, geometric direct minimization (GDM), augmented Roothaan--Hall (ARH), ADIIS, and co-iterative augmented-Hessian (CIAH) methods have improved robustness and convergence \citep{bacskay1981,vanvoorhis2002,host2008,hu2010,sun2017}. More recent trust-region, structured quasi-Newton, model-Hamiltonian, and S-GEK/RVO methods extend this progress \citep{hu2019,helmich2021,sethio2024,qin2024,galvan2025}. The unresolved practical challenge is narrower: can a curvature-aware method beat DIIS in \emph{total wall time} on ordinary molecular calculations, rather than only rescue difficult cases or reduce iteration count?

The cost triangle is simple. First-order history methods are cheap per iteration but may need too many target builds. Exact or nearly exact second-order methods need fewer outer steps, but their Hessian actions can be comparable to additional target Fock builds. Crude diagonal Hessians are cheap but omit the orbital couplings that make Newton directions valuable. CIAH explicitly noted that low-level projected Hessians are possible, but also found that a single-zeta projection could deviate strongly from the target curve in its tests \citep{sun2017}. Structured quasi-Newton methods preserve inexpensive exact Hessian components and approximate expensive components from target-level samples \citep{hu2019}. Model-Hamiltonian methods instead use a cheaper Hamiltonian to improve the wavefunction between exact macroiterations \citep{qin2024}. These ideas point to a missing combination: an independent physical curvature model that is useful from the first step, continuously corrected by exact target-level secants.

\aurora{} implements that combination. It transfers neither the auxiliary energy nor its converged density. The target Hamiltonian supplies every accepted energy, gradient, and convergence decision. The auxiliary model supplies a matrix-free orbital-response metric; transported L-BFGS pairs learn the difference between that metric and the target response. This separation is the central conceptual contribution and the source of the measured wall-time gain.

\section{AURORA: curvature transfer with target-level correction}

Let $C_k$ denote orthonormal molecular orbitals at macroiteration $k$, and let $X$ contain the nonredundant occupied--virtual rotations. A geodesic update is
\begin{equation}
 C_{k+1}=C_k\exp[\kappa(X_k)], \qquad
 \kappa(X)=\begin{pmatrix}0&-X^\dagger\\X&0\end{pmatrix},
 \label{eq:geodesic}
\end{equation}
which preserves orthonormality without a posteriori repair \citep{edelman1998}. The target calculation provides $E_k$, $F_k$, and the exact orbital gradient $g_k$. The base curvature action is written schematically as
\begin{equation}
 A_k[X]=D_k^{\mathrm{gap}}[X]+R_k^{\mathrm{aux}}[X]
          +D_k^{\mathrm{xc}}[X]+\lambda_kD_k[X].
 \label{eq:basecurvature}
\end{equation}
Here $D_k^{\mathrm{gap}}$ retains the target Fock-block energy scale, $R_k^{\mathrm{aux}}$ is the full Coulomb/exchange response evaluated in the valence-STO-3G curvature model, $D_k^{\mathrm{xc}}$ is an inexpensive Kohn--Sham diagonal correction, and the adaptive shift $\lambda_kD_k$ makes the base solve suitable for preconditioned conjugate gradients (PCG).

Accepted target steps produce transported secant pairs
\begin{equation}
 s_k=\mathcal T_k(X_k), \qquad
 y_k=g_{k+1}-\mathcal T_k(g_k),
 \label{eq:secant}
\end{equation}
which correct the low-fidelity curvature through a damped, limited-memory BFGS update \citep{nocedal1980}. The two-loop recursion uses $A_k^{-1}$--computed by matrix-free PCG--as its variable base inverse. The resulting direction is clipped to a trust radius and applied through Eq.~\eqref{eq:geodesic}. A trial is then judged using a new target-level energy and gradient. Rejected trials incur another target evaluation and are included in all reported times and target-build counts.

\begin{figure}[t]
\centering
\includegraphics[width=\textwidth]{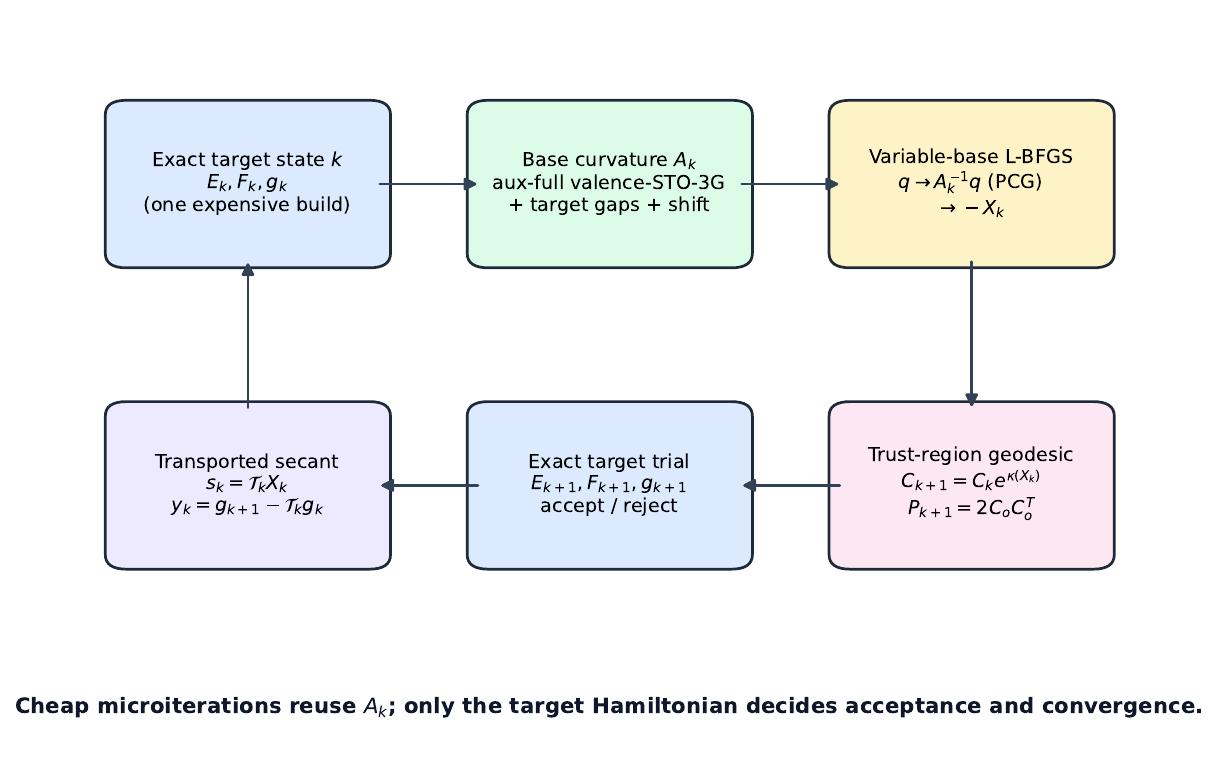}
\caption{\textbf{One \aurora{} macroiteration.} The target Hamiltonian supplies the exact state at the current and trial points. Cheap microiterations reuse an auxiliary full-response curvature operator inside a variable-base L-BFGS/PCG solve. A trust-region geodesic generates a valid density, and the accepted target response becomes a transported secant pair for the next macroiteration.}
\label{fig:macrostep}
\end{figure}

Figure~\ref{fig:macrostep} shows the complete feedback loop. The expensive model is queried for truth; the cheap model supplies full-space curvature; exact secants learn the discrepancy. This differs from solving an auxiliary SCF problem and from transferring an auxiliary density. It also differs from a pure L-BFGS optimizer, which must learn useful curvature only in directions already traversed. Detailed RHF conventions, the auxiliary response, Powell damping, the variable-base recursion, and the exact thin-SVD exponential are derived in the Supplementary Information.

Transported, damped L-BFGS is the default correction because it was stable for both closed- and open-shell calculations. An optional SR1 correction can be faster in some cases, but was less stable, particularly for open shells. Besides the default Coulomb/exchange full-response auxiliary model, the implementation includes ab initio Coulomb-only and diagonal curvature models and semiempirical curvature models. This hierarchy makes the curvature cost tunable rather than fixed.

\section{Results}

\subsection{Direct CPU calculations: the low-overhead DIIS regime}

The cleanest test of the central claim is Supplementary Table~S1: 16 closed-shell, direct RHF pairs on eight molecules with cc-pVDZ and cc-pVTZ basis sets, spanning 137--1484 atomic orbitals. All calculations used the same MINAO initial guess within each pair and a 52-core Intel Xeon Gold 6230R node with 384~GB memory; each calculation used 16 CPU threads. Every \aurora{} calculation used the same Coulomb/exchange full-response valence-STO-3G curvature model.

\begin{figure}[t]
\centering
\includegraphics[width=\textwidth]{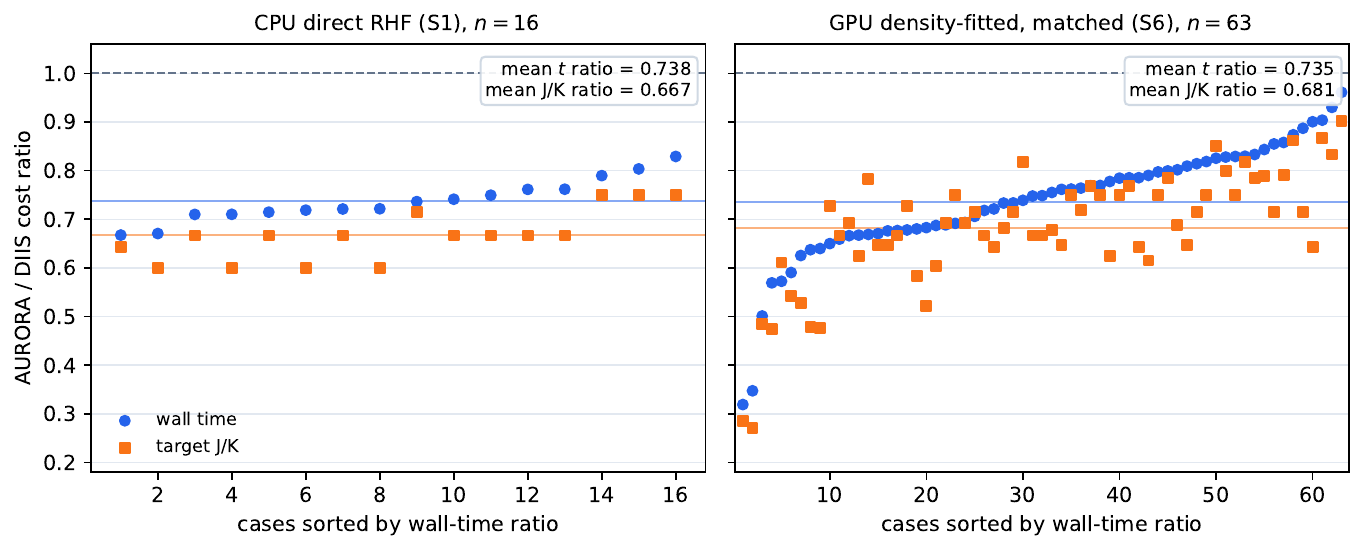}
\caption{\textbf{Paired cost ratios from the prespecified plotting tables.} \textbf{a}, Direct CPU RHF pairs from Supplementary Table~S1. \textbf{b}, Converged, energy-matched density-fitted GPU pairs from Supplementary Table~S6. Each case is sorted by its wall-time ratio; the sorting index is not a molecular-size coordinate. Horizontal solid lines are arithmetic means and the dashed line denotes equal cost. Auxiliary curvature work is included in wall time but not in the target J/K count.}
\label{fig:ratios}
\end{figure}

\aurora{} was faster in all 16 pairs (Fig.~\ref{fig:ratios}a). The mean wall-time ratio $t_\A/t_\D$ was 0.738, corresponding to a 26.2\% reduction, and the mean target J/K ratio was 0.667. The largest absolute final-energy difference was $1.60\times10^{-10}\,\Eh$. The wall-time advantage persisted across the entire size range, rather than appearing only in the largest case.

Focused tests strengthen the mechanism-level interpretation. Across 24 direct CPU pairs that vary molecule, singlet/triplet formalism, and HF/PBE/B3LYP/M06-2X (Table~S2), all 24 were faster; the mean wall-time and target-build ratios were 0.665 and 0.634. A 12-pair direct menthol basis sweep (Table~S3) likewise won every pair, with mean ratios 0.699 and 0.664.

\subsection{Density fitting reveals the overhead boundary}

Density fitting changes the balance because it makes each target J/K build cheaper \citep{dunlap1979,weigend2002}. In the 12-pair density-fitted CPU menthol sweep (Table~S4), the target-build ratio remained 0.664, but the mean wall-time ratio rose to 0.827. 9 of 12 calculations were faster. The three slower calculations were RHF/cc-pVDZ, RHF/cc-pVTZ, and RHF/aug-cc-pVTZ, with wall-time ratios of 1.323, 1.292, and 1.056, respectively. Thus, reducing target iterations is not sufficient when target Fock builds are already very cheap; the auxiliary response and optimizer overhead can dominate. This is an algorithmic crossover, not a contradiction of the curvature mechanism.

The practical policy is therefore conditional. Low-cost density-fitted RHF with a good initial guess should remain on DIIS. As exact exchange, numerical exchange--correlation, direct integrals, basis size, or convergence difficulty raises the cost of a target evaluation, auxiliary curvature becomes increasingly favorable. An automatic production implementation should estimate this ratio during the first few builds and switch solvers accordingly.

The optional SR1 correction shifts this crossover. For density-fitted RHF/cc-pVDZ, RHF/cc-pVTZ, and RHF/aug-cc-pVTZ, its wall times were 1.64, 7.03, and 20.65~s, giving AURORA/DIIS ratios of 1.21, 0.99, and 0.84, respectively. Thus only the smallest basis remained slower than DIIS in this SR1 test, although damped L-BFGS remains the default because of its greater stability across closed- and open-shell calculations.

\subsection{GPU results: broad savings and explicit exclusions}

The direct V100 menthol sweep (Table~S5) contains 12 HF/DFT pairs covering the singlet and both unrestricted and restricted-open-shell triplets. All 12 were faster, with mean wall-time and target-build ratios of 0.674 and 0.663. To reach larger systems at lower cost, the density-fitted V100 set used cc-pVQZ for menthol, cc-pVDZ for vancomycin, and cc-pVTZ for morphine, dibenzo-18-crown-6, beclomethasone, penicillin, and cholesterol. It contains 84 attempted pairs, of which 63 converged to energy-matched endpoints and form Table~S6. The set reaches 1797 atomic orbitals and covers HF, PBE, B3LYP, and M06-2X; open-shell calculations are triplets in unrestricted and restricted-open-shell formalisms.

All 63 included pairs were faster (Fig.~\ref{fig:ratios}b). Their mean wall-time and target-build ratios were 0.735 and 0.681. The largest absolute final-energy difference was $7.06\times10^{-9}\,\Eh$. These results are especially significant because auxiliary curvature work is included in wall time, while the target J/K count records only expensive target-Hamiltonian builds.

The remaining 21 GPU pairs are reported in Table~S7 and excluded from direct speed comparisons. Nine DIIS runs did not converge within 100 cycles; the other pairs ended at energies differing by more than $10^{-4}\,\Eh$. AURORA reached a lower energy in 18 of the 21 cases, and all three cases in which its energy was higher were closed-shell RHF calculations. The open-shell results therefore indicate a marked robustness advantage over DIIS on the difficult cases in this set. Because these pairs do not terminate at matched stationary points, they are reported as convergence outcomes rather than speedup ratios.

\subsection{Why the iteration count falls}

\begin{figure}[t]
\centering
\includegraphics[width=\textwidth]{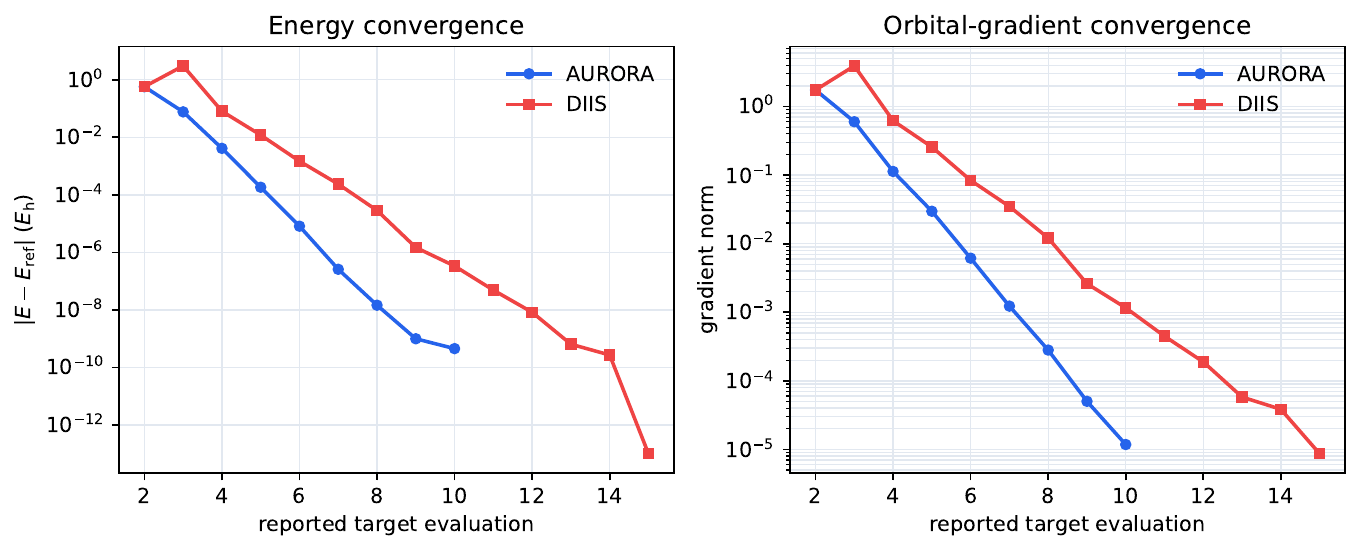}
\caption{\textbf{Representative convergence history for density-fitted B3LYP/cc-pVTZ penicillin on a V100.} Absolute energy error uses the final reported DIIS energy as reference. The first workbook row contains no energy or gradient and is omitted. Data are transcribed from source Tables~9 and 10.}
\label{fig:convergence}
\end{figure}

Figure~\ref{fig:convergence} shows the density-fitted B3LYP/cc-pVTZ convergence history for penicillin on the V100. At target evaluation 9, the AURORA gradient norm is $5.03\times10^{-5}$, compared with $2.61\times10^{-3}$ for DIIS; at evaluation 10 the norms differ by nearly two orders of magnitude. The energy curve shows the same separation. This is the intended Newton-like effect: the auxiliary response coordinates coupled orbital directions before exact target data alone could span them.

In the direct RHF/cc-pVTZ menthol sensitivity test (Table~S8), valence-STO-3G required 9 target builds and 141.84~s, STO-3G required 8 and 129.23~s, and 6-31G required 7 and 115.89~s. Enlarging the curvature basis to cc-pVTZ kept the count at 7 but increased time to 127.12~s. For B3LYP and M06-2X, differences among the tested auxiliary bases were small. Automated selection of the lowest-cost auxiliary model is therefore a natural next step.

\subsection{Summary across benchmark families}

Table~\ref{tab:families} collects the arithmetic mean paired ratios for each benchmark family.

\begin{table}[t]
\centering
\caption{Summary of the benchmark families. Ratios are AURORA/DIIS and values are arithmetic means of the paired ratios.}
\label{tab:families}
\small
\begin{tabularx}{\textwidth}{lrrrr>{\raggedright\arraybackslash}X}
\toprule
Family & $n$ & Faster & Time ratio & Target J/K ratio & Purpose \\
\midrule
S1 CPU direct RHF & 16 & 16 & 0.738 & 0.667 & Molecular size and cc-pVXZ basis scaling \\
S2 CPU direct & 24 & 24 & 0.665 & 0.634 & State formalism and HF/DFT model \\
S3 CPU direct & 12 & 12 & 0.699 & 0.664 & Menthol basis-set sweep \\
S4 CPU DF & 12 & 9 & 0.827 & 0.664 & Cheap-build overhead boundary \\
S5 GPU direct & 12 & 12 & 0.674 & 0.663 & State and model sweep \\
S6 GPU DF, matched & 63 & 63 & 0.735 & 0.681 & Broad energy-matched GPU comparison \\
\bottomrule
\end{tabularx}
\end{table}

\section{Discussion: what constitutes the breakthrough}

The phrase ``beyond the four-decade DIIS default'' is a performance claim, not a claim that no SCF algorithms were developed after 1980. Second-order, direct-minimization, interpolation, surrogate, and model-Hamiltonian methods have made substantial advances. The specific barrier is DIIS's combination of good local convergence and negligible overhead. \aurora{} crosses that barrier in the reported direct CPU, direct GPU, and energy-matched density-fitted GPU families: it converts second-order information into a lower end-to-end time, not just a smaller iteration count.

Three design choices matter. First, the auxiliary model supplies full-space physical coupling from the first step. Second, exact transported secants correct precisely the directions that the target calculation has visited. Third, the target model retains authority over energy, acceptance, and convergence. In compact form,
\begin{equation}
 \boxed{\text{cheap global curvature} + \text{exact local secants}
        + \text{target-only acceptance}.}
\end{equation}
Neither an uncorrected minimal-basis Hessian nor a history-only quasi-Newton model provides the same information pattern.

The present benchmark focuses on main-group organic molecules. Transition-metal systems, near degeneracies, and strong symmetry breaking will be examined in subsequent work, together with automatic auxiliary-model selection.

\section{Conclusion and outlook}

DIIS has remained dominant because it is hard to beat where it matters: total time to an acceptable target-level SCF solution. The present benchmarks show that an independent low-fidelity orbital-response model, corrected by exact target-level L-BFGS secants and constrained by trust-region geometry, can reduce both expensive Fock builds and end-to-end time by roughly 30\% in multiple direct and GPU benchmark families. Across the tested closed- and open-shell HF/DFT calculations, only low-cost density-fitted RHF at small basis size approaches the unfavorable overhead boundary; the optional SR1 correction shifted this boundary so that only cc-pVDZ RHF remained slower in the three-point density-fitted RHF sweep. The practical advance is therefore both an end-to-end acceleration and a measurable criterion for solver selection.

From a general cost perspective, the favorable regime can be written as
\begin{equation}
 C_{\mathrm{exact\ Hessian}}\gg C_{fg}\gg C_{\mathrm{aux}},
 \qquad
 (N_Q-N_A)C_{fg}>N_A\bigl(kC_{\mathrm{aux}}+C_{\mathrm{overhead}}\bigr),
 \label{eq:costcriterion}
\end{equation}
where $C_{fg}$ is the cost of one joint high-fidelity objective/gradient evaluation, $C_{\mathrm{aux}}$ is one auxiliary curvature action, and $N_Q$ and $N_A$ are the macroiteration counts of a first-order/history method and AURORA, respectively. Cost separation alone is insufficient: the auxiliary model must also capture the dominant scales, low-frequency modes, and coupling structure of the target Hessian. AURORA is correspondingly less attractive when the target objective and gradient are cheap, an exact Hessian is readily available, gradients are strongly noisy, or the problem is nonsmooth or dominated by global search.

The nearest extensions remain within electronic structure: HF/KS orbital optimization for larger systems and bases or expensive J/K and exchange--correlation evaluations, followed by orbital-optimized correlated methods, optimized effective potentials, localized-orbital optimization, and density-matrix optimization. CASSCF and MCSCF share the relevant manifold geometry and cost structure, although their much richer stationary-point landscape will require dedicated implementations and benchmarks. Geometry optimization and periodic materials or defect optimization could use semiempirical models, force fields, machine-learned potentials, coarse $k$-point meshes, or lower-level electronic structure models as sources of auxiliary curvature.

More broadly, AURORA suggests a multifidelity structured quasi-Newton framework for expensive scientific computation: retain the high-fidelity objective and acceptance test, transfer only the curvature of a low-cost model, and continually correct it with high-fidelity gradient secants. Promising cost structures include coarse-grid or reduced-order Hessians in PDE-constrained optimization, low-frequency or coarse-propagation models in geophysical inversion, analytic approximate Hessians in computational imaging, and low-accuracy dynamics in optimal control. Extension to scientific machine learning would require variance reduction, periodic full-batch gradients, Fisher or Gauss--Newton auxiliary curvature, and noise-aware trust regions because minibatch noise otherwise degrades secant information.

\section{Implementation and benchmark protocol}

The implementation is built on PySCF 2.8.0 for CPU calculations and GPU4PySCF 1.4.3/CuPy for GPU calculations \citep{sun2018,sun2020,li2025gpu,wu2025gpu}. It supports RHF, UHF, ROHF, RKS, UKS, and ROKS, with direct-SCF and density-fitted target Fock construction. The default correction is transported, damped L-BFGS; SR1 is optional. Available auxiliary curvature models include Coulomb/exchange full response, Coulomb-only and diagonal ab initio forms, and semiempirical forms. The calculations reported in Tables~S1--S8 use the Coulomb/exchange full-response valence-STO-3G option unless an auxiliary-basis sensitivity test is explicitly identified. The AURORA code is available from https://github.com/baopengbp/pyaies/tree/main/scf/aurora.

CPU calculations were run on a 52-core Intel Xeon Gold 6230R node with 384~GB memory, using 16 threads for each calculation. Density-fitted calculations used sufficient memory to keep intermediates in core. GPU benchmarks used one NVIDIA V100 with 32~GB device memory. Molecular geometries were taken from the chain-of-spheres exchange benchmark of Neese and co-workers \citep{neese2009cosx}. 

Within each pair, AURORA and native PySCF DIIS start from the same MINAO guess and use the same molecule, basis, electronic state, density-fitting choice, and convergence settings. State sweeps include the singlet and both unrestricted and restricted-open-shell triplets. The PySCF default energy tolerance, $10^{-9}\,\Eh$, is used; the corresponding default orbital-gradient tolerance is its square root. Wall time includes auxiliary-curvature construction and all rejected target trials. A target J/K count is one requested-Hamiltonian effective-potential build; auxiliary response actions are not counted as target builds. A count of 101 denotes failure to converge within 100 SCF cycles. Native PySCF was used for the direct-SCF DIIS reference calculations.

\section*{Acknowledgements}
This work was supported by the National Natural Science Foundation of China (Grant
number 22173107, 92161205 and 22433006)

\bibliographystyle{plainnat}
\bibliography{references}

\clearpage
\appendix
\part*{Supplementary Information}
\addcontentsline{toc}{part}{Supplementary Information}
\setcounter{section}{0}
\renewcommand{\thesection}{S\arabic{section}}
\setcounter{equation}{0}
\renewcommand{\theequation}{S\arabic{equation}}
\setcounter{table}{0}
\renewcommand{\thetable}{S\arabic{table}}

\section{Gradient directional derivative and RHF Hessian action}

Let the independent occupied--virtual rotations be collected in $\bm\kappa=\{\kappa_{ai}\}$, with energy $E(\bm\kappa)$, gradient $g_{ai}=\partial E/\partial\kappa_{ai}$, and Hessian $H_{ai,bj}=\partial g_{ai}/\partial\kappa_{bj}$. Along $\bm\kappa(t)=\bm\kappa_0+t\bm x$, the multivariate chain rule gives
\begin{equation}
 (H\bm x)_{ai}=\sum_{bj}H_{ai,bj}x_{bj}
 =\left.\frac{d}{dt}g_{ai}(\bm\kappa_0+t\bm x)\right|_{t=0}.
 \label{eq:hvpdir}
\end{equation}
Thus a Hessian--vector product is the directional derivative of the gradient, not a separately defined approximation.

For real closed-shell RHF, partition $C=(C_o\;C_v)$ and take
\begin{equation}
 \kappa(X)=\begin{pmatrix}0&-X^T\\X&0\end{pmatrix}.
\end{equation}
Because $C(t)=C\exp[t\kappa(X)]$,
\begin{equation}
 \delta C_o=C_vX, \qquad \delta C_v=-C_oX^T.
\end{equation}
With the conventional singlet gradient $g=4F_{vo}$ and the unit-occupancy transition density
\begin{equation}
 \Delta D_X=C_vXC_o^T+C_oX^TC_v^T,
\end{equation}
Eq.~\eqref{eq:hvpdir} yields
\begin{equation}
 H^SX=4\left[F_{vv}X-XF_{oo}
 +C_v^T\{2J[\Delta D_X]-K[\Delta D_X]\}C_o\right].
 \label{eq:rhfhvp}
\end{equation}
For canonical orbitals this is equivalent to
\begin{equation}
 H^S_{ai,bj}=4\left[(\epsilon_a-\epsilon_i)\delta_{ab}\delta_{ij}
 +4(ai|bj)-(ab|ij)-(aj|bi)\right].
\end{equation}
The implementation uses the scaled gradient $\tilde g=g/2=2F_{vo}$ and the correspondingly scaled Hessian $\tilde H=H/2$. This scaling leaves the Newton equation and step unchanged and reconciles the factors in the auxiliary-response formula below.

\section{Projected auxiliary full response}

Let $\Pi_k$ denote the implementation's map from the target occupied--virtual tangent space to the independent valence-STO-3G orbital model, and $\Pi_k^*$ its adjoint map back. After density-fitting metric orthogonalization in the auxiliary model, let $\bar B^P_{pq}$ be the three-index tensor in the projected molecular-orbital representation. The closed-shell auxiliary response acting on $X$ is
\begin{align}
 [R_{\mathrm{aux}}(X)]_{ai}
 ={}&8\sum_P \bar B^P_{ai}\sum_{bj}\bar B^P_{bj}X_{bj} \\
 &-2a_x\sum_{Pbj}\left(\bar B^P_{ab}\bar B^P_{ji}
 +\bar B^P_{aj}\bar B^P_{bi}\right)X_{bj},
 \label{eq:auxfull}
\end{align}
where $a_x$ is the exact-exchange fraction. The first line is the Coulomb response; the second retains both exchange contractions, hence \texttt{aux-full}. The target-space action is $\Pi_k^*R_{\mathrm{aux}}(\Pi_kX)$. For the scaled RHF convention, the unshifted base operator is
\begin{equation}
 \mathcal B_k[X]=2(F_{vv}X-XF_{oo})+\Pi_k^*R_{\mathrm{aux}}(\Pi_kX),
 \label{eq:base}
\end{equation}
with the inexpensive Kohn--Sham diagonal term added for DFT. Only the blocks required by Eq.~\eqref{eq:auxfull} are retained. The density-fitting basis used by a target \texttt{density\_fit()} calculation is distinct from this auxiliary orbital-curvature model.

\section{Powell damping and variable-base L-BFGS}

For a stored target secant $(s_i,y_i)$, calculate
\begin{equation}
 b_i=A_ks_i, \qquad \sigma_i=s_i^Tb_i, \qquad \delta_i=s_i^Ty_i,
\end{equation}
where $D_k$ acts as the preconditioner and the shift in $A_k=\mathcal B_k+\lambda_kD_k$ is increased until the base solve has suitable positive curvature. If $\delta_i\ge0.2\sigma_i$, set $\bar y_i=y_i$. Otherwise use Powell damping,
\begin{equation}
 \theta_i=\frac{0.8\sigma_i}{\sigma_i-\delta_i}, \qquad
 \bar y_i=\theta_i y_i+(1-\theta_i)b_i,
\end{equation}
so that $s_i^T\bar y_i=0.2\sigma_i>0$, and define $\rho_i=(s_i^T\bar y_i)^{-1}$. If positive base curvature cannot be established reliably, the pair is discarded rather than forcing an invalid BFGS update.

With recent pairs ordered from oldest to newest, the first L-BFGS loop is
\begin{align}
 q&\leftarrow g_k,\\
 \alpha_i&=\rho_i s_i^Tq, \qquad q\leftarrow q-\alpha_i\bar y_i
 \quad (i=m-1,\ldots,0).
\end{align}
The middle operation is not multiplication by a scalar initial matrix; it is a matrix-free PCG solve,
\begin{equation}
 r\approx A_k^{-1}q.
\end{equation}
The second loop applies the exact-history low-rank correction,
\begin{align}
 \beta_i&=\rho_i\bar y_i^Tr,\qquad
 r\leftarrow r+s_i(\alpha_i-\beta_i)
 \quad(i=0,\ldots,m-1),\\
 X_k&=-r.
\end{align}
Thus the auxiliary model supplies full-space curvature, while target secants repair the visited subspace.

\section{Trust region, exact exponential, and one complete macrostep}

The tangent step is clipped by
\begin{equation}
 X_k\leftarrow \min\left(1,\frac{\Delta_k}{\|X_k\|}\right)X_k.
\end{equation}
For $X=U\Sigma V^T$, the exponential can be evaluated exactly without forming a full molecular-orbital matrix exponential:
\begin{equation}
 \exp[\kappa(X)]=
 \begin{pmatrix}
 I_o+V(\cos\Sigma-I)V^T & -V\sin\Sigma\,U^T\\
 U\sin\Sigma\,V^T & I_v+U(\cos\Sigma-I)U^T
 \end{pmatrix}.
\end{equation}
This thin-SVD expression is algebraically equivalent to the full exponential, not a truncated series.

The core relations for one accepted RHF macrostep are
\begin{equation}
\boxed{\begin{aligned}
 b_i&=A_ks_i,\\
 (s_i,y_i)&\longrightarrow(s_i,\bar y_i,\rho_i),\\
 q&=\text{L-BFGS first loop}(g_k),\\
 r&\approx A_k^{-1}q \quad\text{(PCG)},\\
 X_k&=-\text{L-BFGS second loop}(r),\\
 X_k&\leftarrow\min(1,\Delta_k/\|X_k\|)X_k,\\
 C_{k+1}&=C_k\exp[\kappa(X_k)],\\
 P_{k+1}&=2C_{o,k+1}C_{o,k+1}^T,\\
 (E_{k+1},F_{k+1},g_{k+1})&=\operatorname{TargetSCF}(P_{k+1}),\\
 s_k&=\mathcal T_k(X_k),\\
 y_k&=g_{k+1}-\mathcal T_k(g_k).
\end{aligned}}
\label{eq:macrostep}
\end{equation}
The expensive target calculation appears once for a trial macrostep. If the target energy rejects the trial, the trust radius is reduced and any additional target evaluation is charged to the reported wall time and target-build count. UHF uses separate alpha and beta tangent blocks; ROHF/ROKS uses closed--open, closed--virtual, and open--virtual blocks. Their algebra follows the same operator/transport structure but uses the appropriate generalized Fock blocks.

\section{Supplementary benchmark tables}

In Tables S1--S8, $\A$ and $\D$ denote AURORA and DIIS. Times include the full solver overhead. $N$ counts target-Hamiltonian J/K or effective-potential builds and excludes auxiliary curvature actions. ``T1U'' and ``T1RO'' denote unrestricted and restricted-open-shell triplet formalisms, respectively.

\begin{landscape}
\scriptsize
\setlength{\tabcolsep}{3.2pt}
\begin{longtable}{llrrrrrrrrrr}
\caption{Direct CPU RHF size and basis-set benchmark.}\label{tab:s1}\\
\toprule
Basis & Molecule & $N_{\rm AO}$ & $t_{\rm A}$ (s) & $t_{\rm D}$ (s) & $t_{\rm A}/t_{\rm D}$ & $N_{\rm A}$ & $N_{\rm D}$ & $N_{\rm A}/N_{\rm D}$ & $E_{\rm A}$ ($E_{\rm h}$) & $E_{\rm D}$ ($E_{\rm h}$) & $\Delta E$ ($E_{\rm h}$) \\
\midrule
\endfirsthead
\toprule
Basis & Molecule & $N_{\rm AO}$ & $t_{\rm A}$ (s) & $t_{\rm D}$ (s) & $t_{\rm A}/t_{\rm D}$ & $N_{\rm A}$ & $N_{\rm D}$ & $N_{\rm A}/N_{\rm D}$ & $E_{\rm A}$ ($E_{\rm h}$) & $E_{\rm D}$ ($E_{\rm h}$) & $\Delta E$ ($E_{\rm h}$) \\
\midrule
\endhead
\midrule
\multicolumn{12}{r}{Continued on next page}\\
\endfoot
\bottomrule
\endlastfoot
cc-pVDZ & Cytosine & 137 & 9.70 & 14.53 & 0.668 & 9 & 14 & 0.643 & -392.628646504 & -392.628646504 & 6.03e-12 \\
cc-pVDZ & Histidine & 199 & 13.10 & 18.45 & 0.710 & 10 & 15 & 0.667 & -545.476811082 & -545.476811082 & -3.98e-12 \\
cc-pVDZ & Morphin & 389 & 71.09 & 98.90 & 0.719 & 9 & 15 & 0.600 & -933.862097273 & -933.862097273 & -1.90e-11 \\
cc-pVDZ & Dibenzo-Crown18.6 & 484 & 66.69 & 92.41 & 0.722 & 9 & 15 & 0.600 & -1220.491857867 & -1220.491857867 & 5.00e-11 \\
cc-pVDZ & Beclomethasone & 541 & 164.46 & 215.97 & 0.761 & 10 & 15 & 0.667 & -1683.816586879 & -1683.816586879 & -1.60e-10 \\
cc-pVDZ & Menthol & 254 & 20.29 & 28.39 & 0.715 & 8 & 12 & 0.667 & -465.219330385 & -465.219330385 & 1.10e-11 \\
cc-pVDZ & Penicillin & 430 & 54.95 & 72.13 & 0.762 & 10 & 15 & 0.667 & -1497.163345404 & -1497.163345404 & -2.00e-11 \\
cc-pVDZ & Cholesterole & 622 & 163.37 & 197.03 & 0.829 & 9 & 12 & 0.750 & -1124.150821898 & -1124.150821898 & 1.00e-11 \\
cc-pVTZ & Cytosine & 310 & 20.02 & 27.19 & 0.736 & 10 & 14 & 0.714 & -392.738605242 & -392.738605242 & -3.01e-12 \\
cc-pVTZ & Histidine & 456 & 61.41 & 85.16 & 0.721 & 10 & 15 & 0.667 & -545.630559405 & -545.630559405 & -5.00e-12 \\
cc-pVTZ & Morphin & 896 & 558.38 & 832.55 & 0.671 & 9 & 15 & 0.600 & -934.099283689 & -934.099283689 & -1.60e-11 \\
cc-pVTZ & Dibenzo-Crown18.6 & 1116 & 555.18 & 781.73 & 0.710 & 9 & 15 & 0.600 & -1220.826728159 & -1220.826728159 & 7.00e-11 \\
cc-pVTZ & Beclomethasone & 1250 & 1394.66 & 1860.72 & 0.750 & 10 & 15 & 0.667 & -1684.144911756 & -1684.144911756 & -1.40e-10 \\
cc-pVTZ & Menthol & 610 & 141.56 & 179.25 & 0.790 & 9 & 12 & 0.750 & -465.341816061 & -465.341816061 & -3.01e-12 \\
cc-pVTZ & Penicillin & 976 & 442.13 & 596.47 & 0.741 & 10 & 15 & 0.667 & -1497.471454591 & -1497.471454591 & -3.00e-11 \\
cc-pVTZ & Cholesterole & 1484 & 1456.76 & 1812.89 & 0.804 & 9 & 12 & 0.750 & -1124.428513472 & -1124.428513472 & -2.00e-11 \\
\end{longtable}
\normalsize
\end{landscape}

\begin{landscape}
\scriptsize
\setlength{\tabcolsep}{3.2pt}
\begin{longtable}{lll l r rrrrrrrrr}
\caption{Direct CPU electronic-state and mean-field benchmark.}\label{tab:s2}\\
\toprule
Molecule & State & Basis & Model & $N_{\rm AO}$ & $t_{\rm A}$ (s) & $t_{\rm D}$ (s) & $t_{\rm A}/t_{\rm D}$ & $N_{\rm A}$ & $N_{\rm D}$ & $N_{\rm A}/N_{\rm D}$ & $E_{\rm A}$ ($E_{\rm h}$) & $E_{\rm D}$ ($E_{\rm h}$) & $\Delta E$ ($E_{\rm h}$) \\
\midrule
\endfirsthead
\toprule
Molecule & State & Basis & Model & $N_{\rm AO}$ & $t_{\rm A}$ (s) & $t_{\rm D}$ (s) & $t_{\rm A}/t_{\rm D}$ & $N_{\rm A}$ & $N_{\rm D}$ & $N_{\rm A}/N_{\rm D}$ & $E_{\rm A}$ ($E_{\rm h}$) & $E_{\rm D}$ ($E_{\rm h}$) & $\Delta E$ ($E_{\rm h}$) \\
\midrule
\endhead
\midrule
\multicolumn{14}{r}{Continued on next page}\\
\endfoot
\bottomrule
\endlastfoot
Menthol & S0 & cc-pVDZ & RHF & 254 & 20.29 & 28.39 & 0.715 & 8 & 12 & 0.667 & -465.219330385 & -465.219330385 & 1.10e-11 \\
Menthol & S0 & cc-pVDZ & PBE & 254 & 22.85 & 35.33 & 0.647 & 8 & 13 & 0.615 & -467.715069470 & -467.715069470 & -2.50e-10 \\
Menthol & S0 & cc-pVDZ & B3LYP & 254 & 23.64 & 34.46 & 0.686 & 8 & 12 & 0.667 & -468.353209209 & -468.353209209 & -5.20e-11 \\
Menthol & S0 & cc-pVDZ & M06-2X & 254 & 26.01 & 37.28 & 0.698 & 8 & 12 & 0.667 & -468.142298379 & -468.142298379 & -6.03e-12 \\
Menthol & T1U & cc-pVDZ & RHF & 254 & 41.56 & 59.59 & 0.697 & 15 & 23 & 0.652 & -464.973892115 & -464.973892115 & -1.58e-10 \\
Menthol & T1U & cc-pVDZ & PBE & 254 & 33.92 & 48.47 & 0.700 & 11 & 16 & 0.688 & -467.482156136 & -467.482156136 & -2.09e-10 \\
Menthol & T1U & cc-pVDZ & B3LYP & 254 & 37.46 & 56.75 & 0.660 & 11 & 17 & 0.647 & -468.104000867 & -468.104000866 & -8.60e-10 \\
Menthol & T1U & cc-pVDZ & M06-2X & 254 & 59.45 & 78.06 & 0.762 & 15 & 20 & 0.750 & -467.879962231 & -467.879962231 & -3.67e-10 \\
Menthol & T1RO & cc-pVDZ & RHF & 254 & 41.79 & 59.38 & 0.704 & 15 & 23 & 0.652 & -464.967761583 & -464.967761582 & -1.28e-10 \\
Menthol & T1RO & cc-pVDZ & PBE & 254 & 35.09 & 100.53 & 0.349 & 11 & 31 & 0.355 & -467.481367345 & -467.481367340 & -5.20e-09 \\
Menthol & T1RO & cc-pVDZ & B3LYP & 254 & 38.72 & 70.74 & 0.547 & 11 & 20 & 0.550 & -468.102221036 & -468.102221036 & 2.30e-11 \\
Menthol & T1RO & cc-pVDZ & M06-2X & 254 & 63.65 & 85.24 & 0.747 & 15 & 20 & 0.750 & -467.878052671 & -467.878052671 & 1.80e-11 \\
Cholesterol & S0 & cc-pVDZ & RHF & 622 & 163.37 & 197.03 & 0.829 & 9 & 12 & 0.750 & -1124.150821898 & -1124.150821898 & 1.00e-11 \\
Cholesterol & S0 & cc-pVDZ & PBE & 622 & 212.51 & 332.72 & 0.639 & 9 & 15 & 0.600 & -1130.302429516 & -1130.302429516 & -1.00e-11 \\
Cholesterol & S0 & cc-pVDZ & B3LYP & 622 & 201.60 & 309.43 & 0.652 & 8 & 13 & 0.615 & -1131.841655221 & -1131.841655221 & -1.00e-11 \\
Cholesterol & S0 & cc-pVDZ & M06-2X & 622 & 259.08 & 395.84 & 0.655 & 8 & 13 & 0.615 & -1131.370287472 & -1131.370287472 & 1.98e-11 \\
Cholesterol & T1U & cc-pVDZ & RHF & 622 & 201.37 & 278.14 & 0.724 & 10 & 16 & 0.625 & -1124.024556946 & -1124.024556946 & -1.00e-11 \\
Cholesterol & T1U & cc-pVDZ & PBE & 622 & 282.07 & 466.25 & 0.605 & 10 & 17 & 0.588 & -1130.148279390 & -1130.148279390 & -7.98e-11 \\
Cholesterol & T1U & cc-pVDZ & B3LYP & 622 & 305.91 & 444.01 & 0.689 & 10 & 15 & 0.667 & -1131.685862842 & -1131.685862842 & -1.00e-11 \\
Cholesterol & T1U & cc-pVDZ & M06-2X & 622 & 487.00 & 699.42 & 0.696 & 10 & 15 & 0.667 & -1131.207748868 & -1131.207748868 & -2.00e-11 \\
Cholesterol & T1RO & cc-pVDZ & RHF & 622 & 201.14 & 280.15 & 0.718 & 10 & 16 & 0.625 & -1124.015388008 & -1124.015388008 & -4.00e-11 \\
Cholesterol & T1RO & cc-pVDZ & PBE & 622 & 275.61 & 595.35 & 0.463 & 9 & 20 & 0.450 & -1130.146005429 & -1130.146005427 & -2.72e-09 \\
Cholesterol & T1RO & cc-pVDZ & B3LYP & 622 & 324.11 & 546.64 & 0.593 & 10 & 17 & 0.588 & -1131.683130136 & -1131.683130136 & -1.10e-10 \\
Cholesterol & T1RO & cc-pVDZ & M06-2X & 622 & 533.76 & 674.55 & 0.791 & 10 & 13 & 0.769 & -1131.204837359 & -1131.204837359 & 1.00e-10 \\
\end{longtable}
\normalsize
\end{landscape}

\begin{landscape}
\scriptsize
\setlength{\tabcolsep}{3.2pt}
\begin{longtable}{llrrrrrrrrrr}
\caption{Direct CPU basis-set benchmark for singlet menthol.}\label{tab:s3}\\
\toprule
Basis & Model & $N_{\rm AO}$ & $t_{\rm A}$ (s) & $t_{\rm D}$ (s) & $t_{\rm A}/t_{\rm D}$ & $N_{\rm A}$ & $N_{\rm D}$ & $N_{\rm A}/N_{\rm D}$ & $E_{\rm A}$ ($E_{\rm h}$) & $E_{\rm D}$ ($E_{\rm h}$) & $\Delta E$ ($E_{\rm h}$) \\
\midrule
\endfirsthead
\toprule
Basis & Model & $N_{\rm AO}$ & $t_{\rm A}$ (s) & $t_{\rm D}$ (s) & $t_{\rm A}/t_{\rm D}$ & $N_{\rm A}$ & $N_{\rm D}$ & $N_{\rm A}/N_{\rm D}$ & $E_{\rm A}$ ($E_{\rm h}$) & $E_{\rm D}$ ($E_{\rm h}$) & $\Delta E$ ($E_{\rm h}$) \\
\midrule
\endhead
\midrule
\multicolumn{12}{r}{Continued on next page}\\
\endfoot
\bottomrule
\endlastfoot
cc-pVDZ & RHF & 254 & 20.29 & 28.39 & 0.715 & 8 & 12 & 0.667 & -465.219330385 & -465.219330385 & 1.10e-11 \\
cc-pVDZ & PBE & 254 & 22.85 & 35.33 & 0.647 & 8 & 13 & 0.615 & -467.715069470 & -467.715069470 & -2.50e-10 \\
cc-pVDZ & B3LYP & 254 & 23.64 & 34.46 & 0.686 & 8 & 12 & 0.667 & -468.353209209 & -468.353209209 & -5.20e-11 \\
cc-pVDZ & M06-2X & 254 & 26.01 & 37.28 & 0.698 & 8 & 12 & 0.667 & -468.142298379 & -468.142298379 & -6.03e-12 \\
cc-pVTZ & RHF & 610 & 145.50 & 181.84 & 0.800 & 9 & 12 & 0.750 & -465.341816061 & -465.341816061 & -4.04e-12 \\
cc-pVTZ & PBE & 610 & 121.04 & 188.74 & 0.641 & 8 & 13 & 0.615 & -467.858353271 & -467.858353270 & -3.71e-10 \\
cc-pVTZ & B3LYP & 610 & 150.09 & 197.16 & 0.761 & 8 & 12 & 0.667 & -468.507273261 & -468.507273261 & -1.56e-10 \\
cc-pVTZ & M06-2X & 610 & 146.25 & 209.62 & 0.698 & 8 & 12 & 0.667 & -468.284410914 & -468.284410914 & -1.30e-11 \\
aug-cc-pVTZ & RHF & 966 & 1033.21 & 1361.11 & 0.759 & 9 & 12 & 0.750 & -465.345065402 & -465.345065402 & -1.50e-11 \\
aug-cc-pVTZ & PBE & 966 & 852.33 & 1412.00 & 0.604 & 8 & 14 & 0.571 & -467.863997161 & -467.863997161 & 8.98e-12 \\
aug-cc-pVTZ & B3LYP & 966 & 974.06 & 1417.41 & 0.687 & 8 & 12 & 0.667 & -468.511991846 & -468.511991846 & -2.04e-10 \\
aug-cc-pVTZ & M06-2X & 966 & 1068.73 & 1535.63 & 0.696 & 8 & 12 & 0.667 & -468.289615409 & -468.289615409 & -2.80e-11 \\
\end{longtable}
\normalsize
\end{landscape}

\begin{landscape}
\scriptsize
\setlength{\tabcolsep}{3.2pt}
\begin{longtable}{llrrrrrrrrrr}
\caption{Density-fitted CPU basis-set benchmark for singlet menthol.}\label{tab:s4}\\
\toprule
Basis & Model & $N_{\rm AO}$ & $t_{\rm A}$ (s) & $t_{\rm D}$ (s) & $t_{\rm A}/t_{\rm D}$ & $N_{\rm A}$ & $N_{\rm D}$ & $N_{\rm A}/N_{\rm D}$ & $E_{\rm A}$ ($E_{\rm h}$) & $E_{\rm D}$ ($E_{\rm h}$) & $\Delta E$ ($E_{\rm h}$) \\
\midrule
\endfirsthead
\toprule
Basis & Model & $N_{\rm AO}$ & $t_{\rm A}$ (s) & $t_{\rm D}$ (s) & $t_{\rm A}/t_{\rm D}$ & $N_{\rm A}$ & $N_{\rm D}$ & $N_{\rm A}/N_{\rm D}$ & $E_{\rm A}$ ($E_{\rm h}$) & $E_{\rm D}$ ($E_{\rm h}$) & $\Delta E$ ($E_{\rm h}$) \\
\midrule
\endhead
\midrule
\multicolumn{12}{r}{Continued on next page}\\
\endfoot
\bottomrule
\endlastfoot
cc-pVDZ & RHF & 254 & 1.79 & 1.35 & 1.323 & 8 & 12 & 0.667 & -465.219060647 & -465.219060647 & 1.10e-11 \\
cc-pVDZ & PBE & 254 & 6.06 & 8.90 & 0.681 & 8 & 13 & 0.615 & -467.715468292 & -467.715468291 & -2.53e-10 \\
cc-pVDZ & B3LYP & 254 & 5.22 & 7.41 & 0.705 & 8 & 12 & 0.667 & -468.353485608 & -468.353485608 & -5.40e-11 \\
cc-pVDZ & M06-2X & 254 & 7.59 & 10.81 & 0.702 & 8 & 12 & 0.667 & -468.142305921 & -468.142305921 & -1.00e-11 \\
cc-pVTZ & RHF & 610 & 9.21 & 7.13 & 1.292 & 9 & 12 & 0.750 & -465.341692423 & -465.341692423 & -6.99e-12 \\
cc-pVTZ & PBE & 610 & 16.38 & 24.34 & 0.673 & 8 & 13 & 0.615 & -467.858458372 & -467.858458371 & -3.70e-10 \\
cc-pVTZ & B3LYP & 610 & 15.68 & 20.31 & 0.772 & 8 & 12 & 0.667 & -468.507344165 & -468.507344164 & -1.55e-10 \\
cc-pVTZ & M06-2X & 610 & 25.53 & 35.28 & 0.724 & 8 & 12 & 0.667 & -468.284392287 & -468.284392287 & -2.10e-11 \\
aug-cc-pVTZ & RHF & 966 & 25.83 & 24.46 & 1.056 & 9 & 12 & 0.750 & -465.345009249 & -465.345009249 & -6.99e-12 \\
aug-cc-pVTZ & PBE & 966 & 71.85 & 121.43 & 0.592 & 8 & 14 & 0.571 & -467.864084648 & -467.864084648 & 6.99e-12 \\
aug-cc-pVTZ & B3LYP & 966 & 71.04 & 98.50 & 0.721 & 8 & 12 & 0.667 & -468.512060402 & -468.512060402 & -2.02e-10 \\
aug-cc-pVTZ & M06-2X & 966 & 150.22 & 217.98 & 0.689 & 8 & 12 & 0.667 & -468.289628039 & -468.289628039 & -2.30e-11 \\
\end{longtable}
\normalsize
\end{landscape}

\begin{landscape}
\scriptsize
\setlength{\tabcolsep}{3.2pt}
\begin{longtable}{lll l r rrrrrrrrr}
\caption{Direct V100 GPU state and mean-field benchmark.}\label{tab:s5}\\
\toprule
Molecule & State & Basis & Model & $N_{\rm AO}$ & $t_{\rm A}$ (s) & $t_{\rm D}$ (s) & $t_{\rm A}/t_{\rm D}$ & $N_{\rm A}$ & $N_{\rm D}$ & $N_{\rm A}/N_{\rm D}$ & $E_{\rm A}$ ($E_{\rm h}$) & $E_{\rm D}$ ($E_{\rm h}$) & $\Delta E$ ($E_{\rm h}$) \\
\midrule
\endfirsthead
\toprule
Molecule & State & Basis & Model & $N_{\rm AO}$ & $t_{\rm A}$ (s) & $t_{\rm D}$ (s) & $t_{\rm A}/t_{\rm D}$ & $N_{\rm A}$ & $N_{\rm D}$ & $N_{\rm A}/N_{\rm D}$ & $E_{\rm A}$ ($E_{\rm h}$) & $E_{\rm D}$ ($E_{\rm h}$) & $\Delta E$ ($E_{\rm h}$) \\
\midrule
\endhead
\midrule
\multicolumn{14}{r}{Continued on next page}\\
\endfoot
\bottomrule
\endlastfoot
Menthol & S0 & cc-pVTZ & RHF & 610 & 44.50 & 56.51 & 0.787 & 9 & 11 & 0.818 & -465.341816061 & -465.341816060 & -4.55e-10 \\
Menthol & S0 & cc-pVTZ & PBE & 610 & 8.84 & 13.40 & 0.660 & 8 & 12 & 0.667 & -467.858353271 & -467.858353271 & -2.20e-11 \\
Menthol & S0 & cc-pVTZ & B3LYP & 610 & 40.92 & 56.26 & 0.727 & 8 & 11 & 0.727 & -468.507273261 & -468.507273261 & -1.20e-11 \\
Menthol & S0 & cc-pVTZ & M06-2X & 610 & 42.25 & 57.92 & 0.729 & 8 & 11 & 0.727 & -468.284410914 & -468.284410914 & -2.64e-10 \\
Menthol & T1U & cc-pVTZ & RHF & 610 & 102.13 & 148.06 & 0.690 & 15 & 23 & 0.652 & -465.112952721 & -465.112952721 & -2.15e-10 \\
Menthol & T1U & cc-pVTZ & PBE & 610 & 18.68 & 24.05 & 0.777 & 12 & 16 & 0.750 & -467.636786476 & -467.636786477 & 6.37e-10 \\
Menthol & T1U & cc-pVTZ & B3LYP & 610 & 77.73 & 116.45 & 0.668 & 11 & 17 & 0.647 & -468.269219956 & -468.269219955 & -1.08e-09 \\
Menthol & T1U & cc-pVTZ & M06-2X & 610 & 104.00 & 157.99 & 0.658 & 14 & 22 & 0.636 & -468.032196839 & -468.032196839 & -2.88e-10 \\
Menthol & T1RO & cc-pVTZ & RHF & 610 & 108.13 & 142.11 & 0.761 & 16 & 22 & 0.727 & -465.106187849 & -465.106187848 & -3.53e-10 \\
Menthol & T1RO & cc-pVTZ & PBE & 610 & 18.26 & 55.23 & 0.331 & 12 & 38 & 0.316 & -467.635887374 & -467.635887374 & 4.70e-10 \\
Menthol & T1RO & cc-pVTZ & B3LYP & 610 & 75.99 & 152.31 & 0.499 & 11 & 22 & 0.500 & -468.267255701 & -468.267255701 & -6.99e-12 \\
Menthol & T1RO & cc-pVTZ & M06-2X & 610 & 110.52 & 138.68 & 0.797 & 15 & 19 & 0.789 & -468.029885197 & -468.029885197 & -3.33e-10 \\
\end{longtable}
\normalsize
\end{landscape}

\begin{landscape}
\scriptsize
\setlength{\tabcolsep}{3.2pt}
\begin{longtable}{p{2.6cm}lll r rrrrrrrrr}
\caption{Converged, energy-matched density-fitted V100 GPU pairs.}\label{tab:s6}\\
\toprule
Molecule & State & Basis & Model & $N_{\rm AO}$ & $t_{\rm A}$ (s) & $t_{\rm D}$ (s) & $t_{\rm A}/t_{\rm D}$ & $N_{\rm A}$ & $N_{\rm D}$ & $N_{\rm A}/N_{\rm D}$ & $E_{\rm A}$ ($E_{\rm h}$) & $E_{\rm D}$ ($E_{\rm h}$) & $\Delta E$ ($E_{\rm h}$) \\
\midrule
\endfirsthead
\toprule
Molecule & State & Basis & Model & $N_{\rm AO}$ & $t_{\rm A}$ (s) & $t_{\rm D}$ (s) & $t_{\rm A}/t_{\rm D}$ & $N_{\rm A}$ & $N_{\rm D}$ & $N_{\rm A}/N_{\rm D}$ & $E_{\rm A}$ ($E_{\rm h}$) & $E_{\rm D}$ ($E_{\rm h}$) & $\Delta E$ ($E_{\rm h}$) \\
\midrule
\endhead
\midrule
\multicolumn{14}{r}{Continued on next page}\\
\endfoot
\bottomrule
\endlastfoot
Menthol & S0 & cc-pVQZ & RHF & 1205 & 6.62 & 8.96 & 0.739 & 9 & 11 & 0.818 & -465.371452044 & -465.371452044 & -2.90e-11 \\
Menthol & S0 & cc-pVQZ & PBE & 1205 & 4.68 & 5.92 & 0.790 & 8 & 13 & 0.615 & -467.896912463 & -467.896912463 & 2.98e-10 \\
Menthol & S0 & cc-pVQZ & B3LYP & 1205 & 7.50 & 11.06 & 0.678 & 8 & 11 & 0.727 & -468.543908970 & -468.543908970 & 1.30e-10 \\
Menthol & S0 & cc-pVQZ & M06-2X & 1205 & 9.14 & 14.06 & 0.650 & 8 & 11 & 0.727 & -468.317259865 & -468.317259865 & 2.45e-10 \\
Menthol & T1U & cc-pVQZ & RHF & 1205 & 13.94 & 21.15 & 0.659 & 16 & 24 & 0.667 & -465.151634284 & -465.151634284 & -1.18e-10 \\
Menthol & T1U & cc-pVQZ & PBE & 1205 & 11.55 & 12.78 & 0.904 & 13 & 15 & 0.867 & -467.683309679 & -467.683309679 & -5.08e-10 \\
Menthol & T1U & cc-pVQZ & B3LYP & 1205 & 14.69 & 25.66 & 0.573 & 11 & 18 & 0.611 & -468.313569355 & -468.313569355 & -1.59e-10 \\
Menthol & T1U & cc-pVQZ & M06-2X & 1205 & 27.80 & 40.19 & 0.692 & 15 & 20 & 0.750 & -468.072005967 & -468.072005966 & -9.06e-10 \\
Menthol & T1RO & cc-pVQZ & RHF & 1205 & 13.57 & 20.29 & 0.669 & 18 & 23 & 0.783 & -465.144795807 & -465.144795807 & -2.07e-10 \\
Menthol & T1RO & cc-pVQZ & PBE & 1205 & 12.00 & 34.54 & 0.347 & 13 & 48 & 0.271 & -467.682434872 & -467.682434872 & 6.99e-12 \\
Menthol & T1RO & cc-pVQZ & B3LYP & 1205 & 19.94 & 31.28 & 0.637 & 11 & 23 & 0.478 & -468.311619621 & -468.311619621 & 8.01e-12 \\
Menthol & T1RO & cc-pVQZ & M06-2X & 1205 & 33.31 & 39.49 & 0.843 & 15 & 19 & 0.789 & -468.069384439 & -468.069384439 & -9.40e-11 \\
Morphine & S0 & cc-pVTZ & RHF & 896 & 6.58 & 7.31 & 0.901 & 9 & 14 & 0.643 & -934.098896930 & -934.098896930 & -8.40e-11 \\
Morphine & S0 & cc-pVTZ & PBE & 896 & 4.11 & 5.50 & 0.748 & 10 & 15 & 0.667 & -938.784063994 & -938.784063995 & 4.66e-10 \\
Morphine & S0 & cc-pVTZ & B3LYP & 896 & 6.78 & 8.82 & 0.769 & 10 & 13 & 0.769 & -939.930738461 & -939.930738462 & 2.97e-10 \\
Morphine & S0 & cc-pVTZ & M06-2X & 896 & 7.99 & 11.61 & 0.688 & 9 & 13 & 0.692 & -939.585749157 & -939.585749157 & 4.31e-10 \\
Morphine & T1U & cc-pVTZ & RHF & 896 & 11.88 & 17.46 & 0.680 & 21 & 36 & 0.583 & -933.980150226 & -933.980150226 & -1.01e-10 \\
Morphine & T1U & cc-pVTZ & PBE & 896 & 17.55 & 21.06 & 0.833 & 22 & 28 & 0.786 & -938.649875864 & -938.649875864 & 6.50e-11 \\
Morphine & T1U & cc-pVTZ & B3LYP & 896 & 22.60 & 28.78 & 0.785 & 20 & 26 & 0.769 & -939.790941068 & -939.790941068 & 4.60e-11 \\
Morphine & T1U & cc-pVTZ & M06-2X & 896 & 27.54 & 54.95 & 0.501 & 16 & 33 & 0.485 & -939.435188436 & -939.435188429 & -7.05e-09 \\
Morphine & T1RO & cc-pVTZ & RHF & 896 & 10.12 & 12.26 & 0.826 & 17 & 20 & 0.850 & -933.959961030 & -933.959961030 & -1.61e-10 \\
Morphine & T1RO & cc-pVTZ & B3LYP & 896 & 23.49 & 27.38 & 0.858 & 19 & 24 & 0.792 & -939.788356307 & -939.788356307 & 3.41e-11 \\
Morphine & T1RO & cc-pVTZ & M06-2X & 896 & 28.06 & 38.26 & 0.733 & 15 & 22 & 0.682 & -939.432700603 & -939.432700602 & -2.21e-10 \\
Dibenzo-18-crown-6 & S0 & cc-pVTZ & RHF & 1116 & 8.27 & 11.47 & 0.721 & 9 & 14 & 0.643 & -1220.826159168 & -1220.826159168 & 6.00e-11 \\
Dibenzo-18-crown-6 & S0 & cc-pVTZ & PBE & 1116 & 5.23 & 6.12 & 0.855 & 10 & 14 & 0.714 & -1226.762917908 & -1226.762917909 & 4.50e-10 \\
Dibenzo-18-crown-6 & S0 & cc-pVTZ & B3LYP & 1116 & 9.39 & 13.54 & 0.693 & 9 & 13 & 0.692 & -1228.258323324 & -1228.258323324 & 4.40e-10 \\
Dibenzo-18-crown-6 & S0 & cc-pVTZ & M06-2X & 1116 & 11.01 & 16.52 & 0.666 & 9 & 13 & 0.692 & -1227.773338948 & -1227.773338949 & 5.60e-10 \\
Dibenzo-18-crown-6 & T1U & cc-pVTZ & PBE & 1116 & 35.16 & 36.59 & 0.961 & 37 & 41 & 0.902 & -1226.609495370 & -1226.609495371 & 4.30e-10 \\
Beclomethasone & S0 & cc-pVTZ & RHF & 1250 & 39.15 & 48.08 & 0.814 & 10 & 14 & 0.714 & -1684.144402576 & -1684.144402576 & -3.80e-10 \\
Beclomethasone & S0 & cc-pVTZ & PBE & 1250 & 34.02 & 44.67 & 0.762 & 11 & 17 & 0.647 & -1690.784976230 & -1690.784976230 & 5.70e-10 \\
Beclomethasone & S0 & cc-pVTZ & B3LYP & 1250 & 40.53 & 54.14 & 0.749 & 10 & 15 & 0.667 & -1692.505950496 & -1692.505950497 & 5.20e-10 \\
Beclomethasone & S0 & cc-pVTZ & M06-2X & 1250 & 45.79 & 57.28 & 0.799 & 11 & 14 & 0.786 & -1692.021645252 & -1692.021645252 & 5.00e-10 \\
Beclomethasone & T1U & cc-pVTZ & PBE & 1250 & 75.30 & 91.98 & 0.819 & 24 & 32 & 0.750 & -1690.679482866 & -1690.679482866 & -1.90e-10 \\
Beclomethasone & T1U & cc-pVTZ & B3LYP & 1250 & 84.76 & 143.53 & 0.591 & 19 & 35 & 0.543 & -1692.393731597 & -1692.393731596 & -5.50e-10 \\
Beclomethasone & T1U & cc-pVTZ & M06-2X & 1250 & 112.77 & 143.73 & 0.785 & 21 & 28 & 0.750 & -1691.901988602 & -1691.901988602 & -2.20e-10 \\
Beclomethasone & T1RO & cc-pVTZ & RHF & 1250 & 99.91 & 145.38 & 0.687 & 29 & 48 & 0.604 & -1684.034420047 & -1684.034420046 & -2.80e-10 \\
Beclomethasone & T1RO & cc-pVTZ & M06-2X & 1250 & 120.42 & 159.46 & 0.755 & 21 & 31 & 0.677 & -1691.897562447 & -1691.897562447 & -8.00e-11 \\
Penicillin & S0 & cc-pVTZ & RHF & 976 & 7.51 & 8.47 & 0.887 & 10 & 14 & 0.714 & -1497.470831986 & -1497.470831986 & -1.30e-10 \\
Penicillin & S0 & cc-pVTZ & PBE & 976 & 4.42 & 5.46 & 0.809 & 11 & 17 & 0.647 & -1503.269173069 & -1503.269173069 & 3.70e-10 \\
Penicillin & S0 & cc-pVTZ & B3LYP & 976 & 7.70 & 10.72 & 0.718 & 10 & 15 & 0.667 & -1504.749160194 & -1504.749160195 & 4.60e-10 \\
Penicillin & S0 & cc-pVTZ & M06-2X & 976 & 9.30 & 12.68 & 0.734 & 10 & 14 & 0.714 & -1504.298983930 & -1504.298983930 & 5.20e-10 \\
Penicillin & T1U & cc-pVTZ & PBE & 976 & 21.89 & 68.60 & 0.319 & 28 & 98 & 0.286 & -1503.122502659 & -1503.122502658 & -2.90e-10 \\
Penicillin & T1U & cc-pVTZ & B3LYP & 976 & 31.39 & 35.93 & 0.874 & 25 & 29 & 0.862 & -1504.599180964 & -1504.599180963 & -1.56e-09 \\
Penicillin & T1U & cc-pVTZ & M06-2X & 976 & 38.75 & 57.79 & 0.671 & 22 & 34 & 0.647 & -1504.138440911 & -1504.138440909 & -2.26e-09 \\
Cholesterol & S0 & cc-pVTZ & RHF & 1484 & 50.54 & 60.92 & 0.830 & 9 & 11 & 0.818 & -1124.428218382 & -1124.428218382 & -1.30e-10 \\
Cholesterol & S0 & cc-pVTZ & PBE & 1484 & 41.51 & 52.85 & 0.786 & 9 & 14 & 0.643 & -1130.625847490 & -1130.625847490 & 5.90e-10 \\
Cholesterol & S0 & cc-pVTZ & B3LYP & 1484 & 52.13 & 67.74 & 0.770 & 9 & 12 & 0.750 & -1132.191647036 & -1132.191647037 & 4.20e-10 \\
Cholesterol & S0 & cc-pVTZ & M06-2X & 1484 & 55.53 & 72.86 & 0.762 & 9 & 12 & 0.750 & -1131.689058059 & -1131.689058060 & 5.00e-10 \\
Cholesterol & T1U & cc-pVTZ & RHF & 1484 & 63.02 & 94.42 & 0.667 & 10 & 16 & 0.625 & -1124.303207886 & -1124.303207886 & -6.00e-11 \\
Cholesterol & T1U & cc-pVTZ & PBE & 1484 & 56.88 & 70.94 & 0.802 & 11 & 16 & 0.688 & -1130.473375953 & -1130.473375953 & -6.00e-11 \\
Cholesterol & T1U & cc-pVTZ & B3LYP & 1484 & 71.98 & 101.96 & 0.706 & 10 & 14 & 0.714 & -1132.037121268 & -1132.037121268 & -5.00e-11 \\
Cholesterol & T1U & cc-pVTZ & M06-2X & 1484 & 83.14 & 122.85 & 0.677 & 10 & 15 & 0.667 & -1131.528720564 & -1131.528720563 & -6.00e-11 \\
Cholesterol & T1RO & cc-pVTZ & RHF & 1484 & 66.69 & 98.62 & 0.676 & 11 & 17 & 0.647 & -1124.293237148 & -1124.293237148 & -7.00e-11 \\
Cholesterol & T1RO & cc-pVTZ & PBE & 1484 & 55.72 & 87.08 & 0.640 & 10 & 21 & 0.476 & -1130.470550999 & -1130.470550999 & -5.00e-11 \\
Cholesterol & T1RO & cc-pVTZ & B3LYP & 1484 & 87.17 & 112.09 & 0.778 & 10 & 16 & 0.625 & -1132.033974871 & -1132.033974871 & -1.00e-10 \\
Cholesterol & T1RO & cc-pVTZ & M06-2X & 1484 & 99.15 & 106.61 & 0.930 & 10 & 12 & 0.833 & -1131.525462618 & -1131.525462618 & -9.98e-11 \\
Vancomycin & S0 & cc-pVDZ & RHF & 1797 & 246.40 & 297.19 & 0.829 & 12 & 16 & 0.750 & -5748.961707753 & -5748.961707753 & -1.20e-10 \\
Vancomycin & S0 & cc-pVDZ & PBE & 1797 & 120.02 & 175.76 & 0.683 & 12 & 23 & 0.522 & -5772.956049458 & -5772.956049458 & 8.00e-11 \\
Vancomycin & S0 & cc-pVDZ & B3LYP & 1797 & 249.82 & 313.29 & 0.797 & 12 & 16 & 0.750 & -5778.919042624 & -5778.919042624 & -3.00e-11 \\
Vancomycin & S0 & cc-pVDZ & M06-2X & 1797 & 255.38 & 308.48 & 0.828 & 12 & 15 & 0.800 & -5777.289578683 & -5777.289578683 & -2.00e-11 \\
Vancomycin & T1U & cc-pVDZ & B3LYP & 1797 & 667.08 & 873.05 & 0.764 & 23 & 32 & 0.719 & -5778.787469026 & -5778.787469024 & -2.15e-09 \\
Vancomycin & T1RO & cc-pVDZ & B3LYP & 1797 & 580.36 & 1019.39 & 0.569 & 18 & 38 & 0.474 & -5778.785367643 & -5778.785367643 & 6.09e-10 \\
Vancomycin & T1RO & cc-pVDZ & M06-2X & 1797 & 645.95 & 1032.71 & 0.625 & 19 & 36 & 0.528 & -5777.139849489 & -5777.139849489 & -2.40e-10 \\
\end{longtable}
\normalsize
\end{landscape}

\begin{landscape}
\scriptsize
\setlength{\tabcolsep}{3.2pt}
\begin{longtable}{p{2.6cm}lll r rrrrrrrrr}
\caption{GPU pairs excluded from direct speed comparison because a run reached the 101-build limit or the final energies differed by more than $10^{-4}\,E_{\rm h}$.}\label{tab:s7}\\
\toprule
Molecule & State & Basis & Model & $N_{\rm AO}$ & $t_{\rm A}$ (s) & $t_{\rm D}$ (s) & $t_{\rm A}/t_{\rm D}$ & $N_{\rm A}$ & $N_{\rm D}$ & $N_{\rm A}/N_{\rm D}$ & $E_{\rm A}$ ($E_{\rm h}$) & $E_{\rm D}$ ($E_{\rm h}$) & $\Delta E$ ($E_{\rm h}$) \\
\midrule
\endfirsthead
\toprule
Molecule & State & Basis & Model & $N_{\rm AO}$ & $t_{\rm A}$ (s) & $t_{\rm D}$ (s) & $t_{\rm A}/t_{\rm D}$ & $N_{\rm A}$ & $N_{\rm D}$ & $N_{\rm A}/N_{\rm D}$ & $E_{\rm A}$ ($E_{\rm h}$) & $E_{\rm D}$ ($E_{\rm h}$) & $\Delta E$ ($E_{\rm h}$) \\
\midrule
\endhead
\midrule
\multicolumn{14}{r}{Continued on next page}\\
\endfoot
\bottomrule
\endlastfoot
Morphine & T1RO & cc-pVTZ & PBE & 896 & 19.76 & 72.55 & 0.272 & 22 & 101 & 0.218 & -938.648220107 & -938.644325740 & -3.89e-03 \\
Dibenzo-18-crown-6 & T1U & cc-pVTZ & RHF & 1116 & 32.06 & 26.41 & 1.214 & 33 & 25 & 1.320 & -1220.709669612 & -1220.664315969 & -4.54e-02 \\
Dibenzo-18-crown-6 & T1U & cc-pVTZ & B3LYP & 1116 & 43.58 & 47.02 & 0.927 & 26 & 28 & 0.929 & -1228.103337118 & -1228.092265892 & -1.11e-02 \\
Dibenzo-18-crown-6 & T1U & cc-pVTZ & M06-2X & 1116 & 64.03 & 68.47 & 0.935 & 28 & 30 & 0.933 & -1227.608532034 & -1227.596648293 & -1.19e-02 \\
Dibenzo-18-crown-6 & T1RO & cc-pVTZ & RHF & 1116 & 26.87 & 37.90 & 0.709 & 26 & 39 & 0.667 & -1220.678393041 & -1220.661866955 & -1.65e-02 \\
Dibenzo-18-crown-6 & T1RO & cc-pVTZ & PBE & 1116 & 35.91 & 88.33 & 0.407 & 37 & 101 & 0.366 & -1226.606710639 & -1226.446459621 & -1.60e-01 \\
Dibenzo-18-crown-6 & T1RO & cc-pVTZ & B3LYP & 1116 & 46.68 & 68.00 & 0.686 & 25 & 42 & 0.595 & -1228.098889379 & -1228.095815617 & -3.07e-03 \\
Dibenzo-18-crown-6 & T1RO & cc-pVTZ & M06-2X & 1116 & 69.50 & 66.93 & 1.038 & 28 & 29 & 0.966 & -1227.605517567 & -1227.599364595 & -6.15e-03 \\
Beclomethasone & T1U & cc-pVTZ & RHF & 1250 & 75.45 & 105.37 & 0.716 & 21 & 33 & 0.636 & -1684.037178649 & -1684.061494136 & 2.43e-02 \\
Beclomethasone & T1RO & cc-pVTZ & PBE & 1250 & 88.33 & 264.52 & 0.334 & 29 & 101 & 0.287 & -1690.677480054 & -1690.630598043 & -4.69e-02 \\
Beclomethasone & T1RO & cc-pVTZ & B3LYP & 1250 & 99.15 & 370.96 & 0.267 & 21 & 101 & 0.208 & -1692.389499883 & -1692.358370775 & -3.11e-02 \\
Penicillin & T1U & cc-pVTZ & RHF & 976 & 13.70 & 28.79 & 0.476 & 18 & 44 & 0.409 & -1497.352413090 & -1497.313991278 & -3.84e-02 \\
Penicillin & T1RO & cc-pVTZ & RHF & 976 & 19.37 & 18.70 & 1.036 & 27 & 25 & 1.080 & -1497.283134910 & -1497.328075255 & 4.49e-02 \\
Penicillin & T1RO & cc-pVTZ & PBE & 976 & 26.05 & 71.70 & 0.363 & 32 & 101 & 0.317 & -1503.119744054 & -1503.096693213 & -2.31e-02 \\
Penicillin & T1RO & cc-pVTZ & B3LYP & 976 & 34.72 & 115.27 & 0.301 & 25 & 101 & 0.248 & -1504.594327525 & -1504.572631435 & -2.17e-02 \\
Penicillin & T1RO & cc-pVTZ & M06-2X & 976 & 43.55 & 48.85 & 0.892 & 23 & 28 & 0.821 & -1504.134430424 & -1504.130994407 & -3.44e-03 \\
Vancomycin & T1U & cc-pVDZ & RHF & 1797 & 825.31 & 2225.35 & 0.371 & 32 & 101 & 0.317 & -5748.863668276 & -5748.844011013 & -1.97e-02 \\
Vancomycin & T1U & cc-pVDZ & PBE & 1797 & 317.67 & 808.02 & 0.393 & 32 & 101 & 0.317 & -5772.836575995 & -5772.624304712 & -2.12e-01 \\
Vancomycin & T1U & cc-pVDZ & M06-2X & 1797 & 617.49 & 1052.26 & 0.587 & 20 & 37 & 0.541 & -5777.142384946 & -5777.141648327 & -7.37e-04 \\
Vancomycin & T1RO & cc-pVDZ & RHF & 1797 & 803.40 & 805.66 & 0.997 & 30 & 33 & 0.909 & -5748.814032938 & -5748.814900986 & 8.68e-04 \\
Vancomycin & T1RO & cc-pVDZ & PBE & 1797 & 298.15 & 810.58 & 0.368 & 30 & 101 & 0.297 & -5772.835788005 & -5771.690334390 & -1.15e+00 \\
\end{longtable}
\normalsize
\end{landscape}

\begin{landscape}
\scriptsize
\setlength{\tabcolsep}{3.2pt}
\renewcommand{\arraystretch}{0.86}
\begin{longtable}{ll lrr}
\caption{Auxiliary-curvature basis sensitivity for menthol on CPU.}\label{tab:s8}\\
\toprule
Target basis & Target calculation & Curvature basis & Wall time (s) & Target J/K builds \\
\midrule
\endfirsthead
\toprule
Target basis & Target calculation & Curvature basis & Wall time (s) & Target J/K builds \\
\midrule
\endhead
\midrule
\multicolumn{5}{r}{Continued on next page}\\
\endfoot
\bottomrule
\endlastfoot
ccpvdz & rhf aux-full 1G  & valence-STO-3G & 20.05 & 8 \\
ccpvdz & rhf aux-full 1G  & STO-3G & 20.02 & 8 \\
ccpvdz & rhf aux-full 1G  & 6-31G & 17.93 & 7 \\
ccpvdz & rhf aux-full 1G  & 6-31G* & 18.53 & 7 \\
ccpvdz & rhf aux-full 1G  & cc-pVDZ & 18.96 & 7 \\
ccpvtz & rhf aux-full df 100G  & valence-STO-3G & 8.93 & 9 \\
ccpvtz & rhf aux-full df 100G  & STO-3G & 8.58 & 8 \\
ccpvtz & rhf aux-full df 100G  & 6-31G & 9.62 & 7 \\
ccpvtz & rhf aux-full df 100G  & 6-31G* & 11.47 & 7 \\
ccpvtz & rhf aux-full df 100G  & cc-pVDZ & 13.30 & 7 \\
ccpvtz & rhf aux-full df 100G  & cc-pVTZ & 25.63 & 7 \\
ccpvtz & rhf aux-full 1G  & valence-STO-3G & 141.84 & 9 \\
ccpvtz & rhf aux-full 1G  & STO-3G & 129.23 & 8 \\
ccpvtz & rhf aux-full 1G  & 6-31G & 115.89 & 7 \\
ccpvtz & rhf aux-full 1G  & 6-31G* & 115.93 & 7 \\
ccpvtz & rhf aux-full 1G  & cc-pVDZ & 117.06 & 7 \\
ccpvtz & rhf aux-full 1G  & cc-pVTZ & 127.12 & 7 \\
ccpvdz & rb3lyp aux-full 1G  & valence-STO-3G & 22.68 & 8 \\
ccpvdz & rb3lyp aux-full 1G  & STO-3G & 22.47 & 8 \\
ccpvdz & rb3lyp aux-full 1G  & 6-31G & 22.60 & 8 \\
ccpvdz & rb3lyp aux-full 1G  & 6-31G* & 22.68 & 8 \\
ccpvdz & rb3lyp aux-full 1G  & cc-pVDZ & 22.76 & 8 \\
ccpvtz & rb3lyp aux-full df 100G  & valence-STO-3G & 14.91 & 8 \\
ccpvtz & rb3lyp aux-full df 100G  & STO-3G & 14.89 & 8 \\
ccpvtz & rb3lyp aux-full df 100G  & 6-31G & 15.22 & 8 \\
ccpvtz & rb3lyp aux-full df 100G  & 6-31G* & 15.60 & 8 \\
ccpvtz & rb3lyp aux-full df 100G  & cc-pVDZ & 15.87 & 8 \\
ccpvtz & rb3lyp aux-full df 100G  & cc-pVTZ & 18.16 & 8 \\
ccpvtz & rb3lyp aux-full 1G  & valence-STO-3G & 136.23 & 8 \\
ccpvtz & rb3lyp aux-full 1G  & STO-3G & 133.95 & 8 \\
ccpvtz & rb3lyp aux-full 1G  & 6-31G & 137.30 & 8 \\
ccpvtz & rb3lyp aux-full 1G  & 6-31G* & 144.47 & 8 \\
ccpvtz & rb3lyp aux-full 1G  & cc-pVDZ & 145.11 & 8 \\
ccpvtz & rb3lyp aux-full 1G  & cc-pVTZ & 147.52 & 8 \\
ccpvdz & rm062x aux-full 1G  & valence-STO-3G & 58.38 & 15 \\
ccpvdz & rm062x aux-full 1G  & STO-3G & 58.50 & 15 \\
ccpvdz & rm062x aux-full 1G  & 6-31G & 58.51 & 15 \\
ccpvdz & rm062x aux-full 1G  & 6-31G* & 58.42 & 15 \\
ccpvdz & rm062x aux-full 1G  & cc-pVDZ & 58.73 & 15 \\
\end{longtable}
\normalsize
\end{landscape}

\end{document}